# A QM/MD coupling method to model the ion-induced polarization of graphene

Joshua D Elliott[1], Alessandro Troisi[2], Paola Carbone[1]

[1]*Department of Chemical Engineering and Analytical Science, University of Manchester, Manchester M13 9PL United Kingdom*
[2]*Department of Chemistry, University of Liverpool, Liverpool L69 7ZD United Kingdom*

*Email:* joshua.elliott@manchester.ac.uk, paola.carbone@manchester.ac.uk

**Abstract**

We report a new Quantum Mechanical/Molecular Dynamics (QM/MD) simulation loop to model the coupling between the electron and atom dynamics in solid/liquid interfacial systems. The method can describe simultaneously both the quantum mechanical surface polarizability emerging from the proximity to the electrolyte, and the electrolyte structure and dynamics. In the current set up Density Functional Tight Binding calculations for the electronic structure calculations of the surface are coupled with classical molecular dynamics to simulate the electrolyte solution. The reduced computational cost of the QM part makes the coupling with a classical simulation engine computationally feasible and allows simulation of large systems for hundreds of nanoseconds. We tested the method by simulating both a non-charged graphene flake and a non-charged and charged infinite graphene sheet immersed in an NaCl electrolyte solution. We found that, when no bias is applied, ions preferentially remained in solution and only cations are mildly attracted to the surface of the graphene. This preferential adsorption of cations *vs* anions seems to persist also when the surface is moderately charged and rules out any substantial ions/surface charge transfer.

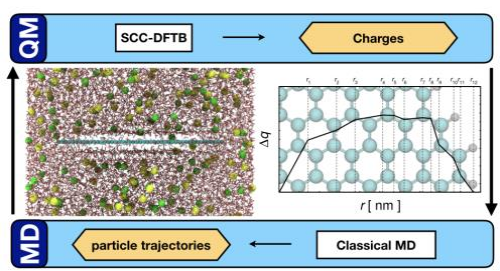

## 1. Introduction

Graphene-based supercapacitors[1–6] (electrical double layer capacitors) are an emergent technology capable of energy storage and charge/discharge rates at levels that are orders of magnitude larger than conventional capacitors and batteries.[7] Their function leverages on the high relative surface area that accompanies low dimensional materials and on an electrostatic charge storage mechanism, which is based on the physisorption of ionic species at the surface.[5,8–10] When the electrolyte is an aqueous solutions, experimental and theoretical investigations have proven that the kinetics and thermodynamics of the physisorption process are the result of a delicate balance between hydration free energy and surface effects. However, results can often appear contradictory; even small (atomic-scale) defects in the structure of the graphitic surface, its geometry, dimensionality and chemical modifications may extensively affect the experimental measurements. For example Yang *et al* report that the basal capacitance of graphene films made from graphene oxide is independent of the nature of the adsorbing cations,[11] however similar electrochemical measurements, performed on activated carbon, indicate that the capacitance is to some degree ion-specific ($Li_+$ less adsorbed than $Na_+$ and $K_+$).[12] In contrast, experiments performed on single walled carbon nanotubes (SWCNT) suggest that $Li_+$ should have higher affinity to the surface than $Na_+$ and $K_+$,[13] and moreover also show that the nature of the counterions can also affect the results. Recently, Iamprasertkun *et al,* using Highly Ordered Pyrolitic Graphite (HOPG) as proxy for the graphene surface, found that the HOPG basal capacitance is ion-dependent with a trend that follows the ionic size and hydration free energy.[14]

Rationalizing these ostensibly conflicted experimental results is a difficult task, which has led to an extensive catalogue of molecular simulations that attempt explain and predict ions adsorption at these interfaces. Classical molecular simulations have shown the importance of including a detailed atomistic description of the interface, which accounts for both the interactions within the first solvation shell and for long-range effects.[15] In particular, ions and water polarization, solvent exclusion, and hydrogen bonding rearrangements determine whether a particular ion is found at aqueous interfaces.[16–18] Despite these promising results, classical models often fail when simulating solid/liquid interfaces where polarizability effects play an important role, for instance in the modulation of the Coulomb interactions between electrolyte solutions and graphitic surfaces. In order to capture these important phenomena, molecular models employed to simulate interfacial systems need to include the polarizability of all the species involved. In the majority of classical simulations, however, only (if any) the polarizability of the electrolyte is included, assuming that the surface polarization has a negligible effect on water structuring and dynamics at the interface.[19] In the case of metallic/semimetallic surfaces, such as graphene or carbon nanotubes (CNT), which have an abundance of aromatic rings with delocalized $\pi$-electrons, this assumption is questionable. Recent Density Functional Theory (DFT) calculations have showed that small monovalent metal ions are able to significantly polarize carbonaceous nanostructures such as CNT[20] and graphene quantum dots.[21] In the latter case, the polarization effect is so important that the band-gap of the finite graphene sheet is also modified. The importance of surface polarization has recently been highlighted by other first-principles molecular dynamics (FPMD) simulations that showed that cations such as $Na_+$ preferentially reside at the interface when confined in a CNT.[22] This is in contrast to their behaviour at unstructured, non-polarizable surfaces, in which cations are repelled. Other simulations have also shown that by only incorporating the polarizability of the ions, molecular simulations predict the wrong adsorption tendency between $Na_+$ and $K_+$ on the surface of a CNT.[23]

First principles-based approaches are the only computational methods that can be used to capture the fluctuation of surface charges which is brought about by the presence of the electrolyte solutions. However due to their comparably high computational cost, FPMD simulations are typically restricted to several hundred water molecules, unrealistic low ionic concentrations and short dynamical trajectories on the order of hundreds of picoseconds.[24–26] Classical polarizable force fields for the surface and the electrolyte can be used to circumvent the problem of the conventional fixed-charge

potential.27 However, because of the semimetallic and 2D nature of graphene it is very difficult to know *a-priori* whether a model with atomic-centered polarizability can capture the polarization of graphene and a case-by-case parameterization might be needed. Recently we proposed a DFT free-energy optimized Lennard-Jones type potential to account of the polarization of all the system species.14 The advantage of this approach is that the simulations are not slowed down by the inclusion of extra dummy charge points as in polarizable models and, since we used the implicit solvation model during the DFT calculations, both hydration and polarization effects were considered.28,29 This model however has also shortcomings: it was parameterised using a single ion so the effect of ionic screening due to multiple ions was only included via the (untested) standard Lorentz-Berthelot combination rules, and the model did not account for the polarization of the graphene surface due to the specific arrangement of the water molecules at the interface.30,31

Finally, since in almost all applications and electrochemical experiments the graphitic surface is electrified, it is desirable to have a computational approach that is able in principle to describe the polarizability, while simultaneously accommodating for the presence of a net surface charge. Recently, Zhan *et al* used a fully integrated first principle/continuum model to investigate the adsorption of cations on electrified graphitic surfaces.32 The method accounts for the electronic structure of the surface and of a single approaching cation, treating these with DFT, and describes the rest of the electrolyte solution using the RISM approach (Reference Interaction Site Model)33 These simulations, which can account also for the screening effect due to the ionic strength of the solution, show surface attraction for almost all cations investigated and indicated, in agreement with the experimental data of Iamprasertkun *et al*,34 that the adsorption energy increases with the ionic radius. As presented, the method is yet to describe any of the dynamical aspects associated with ion adsorption process and also does not account explicitly for neither the water molecules and their rearrangement at the surface nor for the other ions in the electrolyte solutions.

There is therefore the need to develop a new method that can capture simultaneously the structure and dynamics of the interface accounting for both the thermodynamics of the solution and the electronic polarization of the surface. Here we propose a novel Quantum Mechanical/ Classical Molecular Dynamics (QM/MD) approach that, taking advantage of the specific electronic structure of graphene, couples the semi-empirical calculation of the surface charge with the *classical* dynamics of the electrolyte. This is achieved creating a workflow where density functional tight binding (DFTB) calculations of the graphene atomic charges are nested in a loop of molecular dynamics simulations that allow the configuration of the electrolyte to evolve over time in response to the change in the graphene polarization. In what follows we present the core idea of the coupling and apply the QM/MD procedure to the test case of an uncharged graphene sheet immersed in NaCl electrolyte solutions of different ionic strengths with the aim of quantifying the surface polarization and how it contributes the ion-adsorption mechanism.

## 2. Method

In this work, we developed a novel QM/MD workflow to introduce within a classical model, the electrolyte-induced polarization of the graphene flake. The procedure takes into account the dependence of the surface polarization on the electrolyte configuration recalculating the partial charges associated to each carbon atom every $\tau$ ps of a molecular dynamics simulation, during which the atoms in the electrolyte solution are allowed to move.

Electronic structure calculations of the graphene atoms are carried out using the self-consistent charge density functional tight binding, SCC-DFTB, approach, which is an approximation to Kohn-Sham density functional theory.[35] We opted for SCC-DFTB over other electronic structure methods owing to its favourable accuracy-computational viability tradeoffs.[36] The DFTB method offers extremely rapid solutions of the electronic problem since it utilizes parameterized Hamiltonian matrix elements. Moreover, the second order, SCC approximation captures redistribution of electronic density in the form of atomic charges. For the interested reader, there are several instances where this method has been recounted in great detail.[35,37]

When coupled to a classical atomistic environment, which includes water and ions, SCC-DFTB has proven very successful in describing charge transport phenomenon in biomolecular systems.[38–43] These examples share striking similarities with the surface-electrolyte interfaces present in graphene-based supercapacitors. Yet, an important difference is that, in the most common applications of SCC-DFTB/MM simulations, the primary interest is the dynamical behaviour of the quantum mechanically embedded region, with the classical part providing the aqueous background environment. Instead, in the study of non-faradic ion-surface interactions we are concerned with the structuring and dynamical evolution of the classical components of the simulation, the ionic species and the water molecules. Consequently, these QM/MD simulations are most reliant on the ability of SCC-DFTB to reproduce the correct electronic band structure and static polarizability of graphene,[44] and the striking features of 0 dimensional graphene flakes that arise due to quantum confinement effects.[45–47]

The computational workflow makes iterative calls between Molecular Dynamics (MD) and SCC-DFTB software as shown schematically in Figure 1. More specifically, in the MD to QM step the coordinates of the electrolyte, i.e. those associated with ions and solvent molecules, are converted into a set of point charges $\{Q_i^{\text{sol}}\}$. The set $\{Q_i^{\text{sol}}\}$ then forms a background electrostatic potential for the quantum mechanical calculations. For anions and cations the values of $Q_i$ are chosen according to the formal charges associated with the specific ions involved (for example +/− 1 for monovalent ions), while for the water molecules the partial charges associated to the oxygen and hydrogen atoms are taken from the chosen water model (in this case SCP/E[48] see more below).

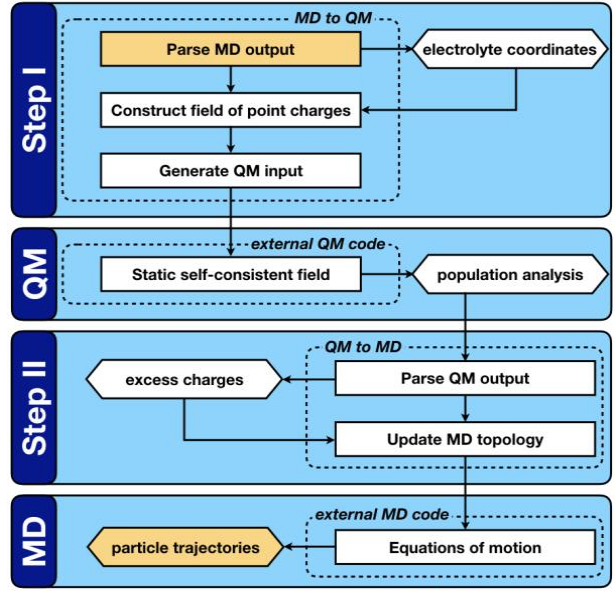

**Figure 1:** Schematic representation of the QM/MD workflow. Key computable quantities are represented by hexagonal boxes and square boxes represent computational processes. The two boxes coloured gold link sequential iterations.

In the QM step the quantum mechanically derived charges on the carbon and hydrogen atoms in the graphene flakes are extracted from the SCC-DFTB output data and used to update the force-field for the MD simulation. The graphene charges are obtained from a Mulliken population analysis[49] and converted into an excess-charge representation, $\{\tilde{q}_i\}$, the details of which are provided below. Whilst we recognize that Mulliken population analysis can be highly sensitive to the adopted basis set, in our case this charge partitioning scheme has the distinct advantage of ensuring the full equivalence between the DFTB and classical forces acting on the electrolyte atoms,[50] we verified this numerically below.

Due to the finite size of the graphene flake, the large permanent polarization of the C-H bonds at the graphene flake edges give rise to the formation of a molecular quadrupole moment in the plane of the graphene flake. This has a disproportionately strong role in the ensuing ion adsorption behaviour,[51–53] resulting in the over-binding of cations and under-binding of anions when compared with semi-infinite graphene models.[27] In order to remove this edge effect, we define a set of so-called excess charges that are computed in between the QM and MD steps. Let $q_i^{\text{vacuum}}$ be the computed Mulliken charge on an atom $i$ in the pristine graphene flake in vacuum, and $q_i$ be the Mulliken charge on the same atom computed from one of the solvated snapshots during the QM/MD loop. Then we define the excess charge $\tilde{q}_i$ as

$$\tilde{q}_i = q_i - q_i^{\text{vacuum}} \qquad \qquad 1$$

This set of excess charges is constructed such that the permanent C-H bond dipoles are neutralized, while the polarization brought about by the point charges is maintained since $q_i^{\text{vacuum}} \approx 0$ for bulk atoms and $q_i \approx q_i^{\text{vacuum}}$ for edge atoms.

The system, described by the new set of carbon partial charges, $\{\tilde{q}_i\}$, is propagated using classical molecular dynamics, and the recalculation of the $\{\tilde{q}_i\}$ is performed at fixed time intervals, $\tau$. Whilst the overall length of the QM/MD simulation is subject to the usual atomistic MD considerations (equilibration of energy, temperature and pressures), the choice of $\tau$ is more delicate and discussed in more detail below.

### 3. Computational Details

In order to investigate the behaviour of ions at the graphene-electrolyte interface we use our QM/MD strategy on two different set ups. The first is a system comprised of a graphene flake solvated in NaCl solutions. The simulation box has dimensions $7.21 \times 7.87 \times 10.00$ nm and is filled with fully dissociated NaCl electrolyte solutions at 3 different ionic concentrations: 0.0 (e.g. single ion), 0.5 and 1.0 M. The graphene flake has an overall diameter of 5nm. The positions of the atoms in the graphene flake are kept fixed during MD simulations. Figure 3 shows snapshots on the simulation box. In the second set-up we use a $2.97 \times 3.00 \times 16.00$ nm$^3$ cell which contains one semi-infinite graphene sheet, containing 336 C atoms, periodic in the *x* and *y* directions. The sheet is in contact with a slab of 1M aqueous NaCl, 8 nm thick, and a further 8 nm of vacuum separates the aqueous phase with the repeated periodic image (see Figure 7a). With this set-up we performed simulations with non-charged and charged graphene, the overall charge applied was +/− 4*e*, which corresponds to a net surface charge density of $\sigma_S$ = +/− 0.44*e* nm$^{-2}$, this is equivalent to previous classical investigations of graphene supercapcitors.[54] In accordance, the electrolyte composition varies slightly in order to keep the overall charge of the cell zero. All the results presented below will refer to the first set up unless it is said otherwise.

Molecular dynamics calculations in the NVT ensemble are carried out using the GROMACS code,[55,56] version 2018.4. A time step of 1 fs and the leapfrog algorithm are used to integrate the equations of motion at a constant temperature of 298.15 K, which is kept constant using the Nosé-Hoover thermostat ($\tau_T = 0.1$ ps). Long-range electrostatic interactions are treated using the reaction field approach, with a cut-off of 1.4 nm. Non-bonded interactions are computed using a Lennard-Jones 12-6 potential, which is truncated smoothly at 1.2 nm using a switch function starting at a distance of 1.0 nm.

In all simulations the graphene atoms are frozen, for the intermolecular parameters, we have used the following parameters: Water in our simulations is modelling using the SPC/E model[48] with the SETTLE algorithm used to maintain rigid molecule geometries.[57] This is compatible with the Werder water-graphene parameters that give rise to the experimentally measured water contact angle.[15] Cl and Na ion parameters, also compatible with the SPC/E model, are taken from the work of Cheatham *et al*.[59]

To avoid complications associated with interacting point charges subject to periodic boundary conditions, single-point SCC-DFTB calculations have been carried out in open-boundary conditions using the DFTB+ code.[36] In the volume occupied by the simulation box, the continuum is described by point charges, representative of the classical water molecules and ions. The coordinates of the graphene flake are fixed, C-C (1.427 Å) and C-H (1.089 Å) bond lengths were optimized at the periodic PBE-DFT level. The empirical description of the interactions between C and H atoms are provided by the mio-1-1 parameter set.[35]

In order to minimize the time associated to the QM calculations, both the choice of the SCC threshold and the Fermi temperature have been optimized, carrying out several DFTB test calculations using different values of these simulation parameters. It was observed (see supporting information Figure S1) that both parameters have a sizeable effect on the simulation time and relatively minimal impact on the values of the atomic charges. Considering that the charges are used as data in a classical force field, an error of less then 1% was deemed acceptable. Therefore, unless otherwise stated explicitly, the SCC threshold was always set to $1 \times 10^{-2}$ Hartree and a Fermi temperature to 300 K. In each SCC-DFTB step the orbitally resolved charges for each graphene atom are initialized from optimized charges from the DFTB previous step.

Finally, to choose the value of the time interval, $\tau$, the average change atomic charges, $\overline{\Delta q}$, was calculated for $\tau = 2$ fs (i.e. every MD time step) and at increasing intervals up to $\tau = 7$ ps (see supporting information Figure S2). The result ($\overline{\Delta q}$ =0.011 $e$ and 0.015 $e$ for $\tau = 2$ fs and $\tau = 5$ ps respectively) indicates that a choice of $\tau = 5$ ps represents a good compromise between an accurate representation of the change in polarization and overall simulation time.

## 4. Results

### *A. Calculation of the graphene polarization from DFTB*

The QM/MD procedure allows us to quantify the graphene flake polarization induced by the proximity of the ions. To do this we calculate the charge distribution of a C96 graphene flake simulated at the DFT and DFTB levels of theory with a point charge ($q = -1.0$ $e$) placed directly above the centre of the flake at a distances, $d_\perp$, varying between 0.3 and 0.6 nm, as depicted in Figure 2(a). The comparison between the two levels of theory allows also to verify the capability of the DFTB method to correctly reproduce the graphene polarizability using the point-charge description of the ions.

In these tests the DFT calculations are carried out using the Gaussian 09 software distribution; electronic wavefunctions are expanded in the 6-31G basis set[60–62] with the three parameter, hybrid B3LYP functional[63] used to describe electron exchange and correlation interactions. This computational setup has been selected over other possible basis set XC potential combinations since it matches the original DFTB mio-1-1 Hamiltonian parameterization.[35] In addition, the use of a smaller basis set allows us to avoid known complications associated with the use of diffuse orbitals in the modelling of highly symmetric aromatic systems.[64]

To map the redistribution of charge density in response to the proximity of the point charge, in Figure 2(b) we plot the integrated total atomic charge difference $\Delta q_{\text{Tot}}$

$$\Delta q_{\text{Tot}}(\mathbf{r}) = \int_0^{|\mathbf{r}|} \left[\sum_i \tilde{q}_i(\mathbf{r})\right] d\mathbf{r} \qquad 2$$

as a function of the radial distance from the centre of the flake. Here **r** is the vector which describes the position of an atom relative to the centre of the flake and $\tilde{q}_i$ is the computed charge difference on atom *i* between the polarized and unpolarized graphene flake as described by Equation 1. This analysis captures the accumulation and depletion of charge density within each atomic ring of the graphene flake. In response to the

(negative) point charge, we observe positive charge accumulates in the innermost atomic rings $r_1$ to $r_3$. In the buffer region, which is defined from $r_4$ to $r_8$, the presence of the ion is less strongly felt and changes to the polarization are considerably smaller. Finally, to compensate for the accumulation of charge density close to the point charge, positive charge density is lost from the outer regions $r_9$ to $r_{12}$. As anticipated, the polarization of the flake increases as the point charge is brought closer to the flake, a trend captured by both DFT and DFTB. Our results indicate that B3LYP describe a slightly more polarizable graphene flake than DFTB (5-8% more charge is displaced by the periphery to the centre of the flake). However, given than B3LYP can overestimate the static polarizability by up to 4% with respect to higher levels of theory,[65] we deem the DFTB to be acceptable for the QM/MD simulations carried out here.

Figure 2(b) clearly indicates that the surface charge redistribution induced by the proximity of the ion is sizable and cannot be neglected during a simulation. Figure 2(b) allows also to appreciate the non-local nature of the charge redistribution which affects carbon atoms as far as 3 bonds suggesting that atomic-centered polarizable models can fall short in capturing the full physics of the physisorption process.

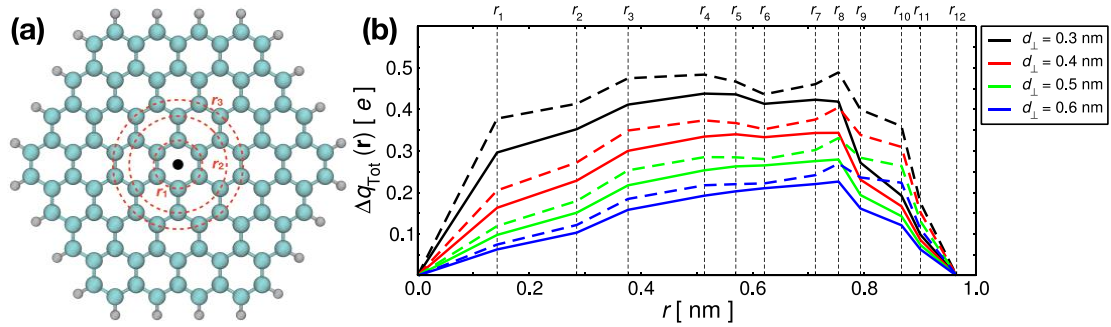

**Figure 2:** (a) Geometry of the hexagonal C96 graphene flake; dashed red circles illustrate radially equivalent atoms at increasing distances $r$ from the location of the point charge (black dot). (b) Plot comparing the DFTB (solid) and DFT (dashed) integrated Mulliken charges as a function of radius $r$ and point-charge adsorption height $d_\perp$ ($q = -1.0\ e$). Vertical dashed lines mark the radii of the C atoms in the C96 flake.

Our method neglects the possible charge-transfer that may occur between ions and surface during adsorption.[32] This is a short-range interaction which, at the reported adsorption heights relevant for solvated cations and anions ($\approx$ 3 Å see Figure 4 and 5 below), accounts for less than 0.04 $e$ (Li+, Na+, K+) becoming therefore negligible compared to the magnitude of the overall Coulomb and Lennard-Jones non-bonded interactions. In case charge transfer becomes a relevant phenomenon for the physisorption process, this term, being very short range, could be anyway incorporated into the short range non-bonded interaction.

*B. Numerical Equivalence of Coulomb Forces based on Mulliken Charges*

In order to verify full consistency in the calculation of the Coulomb interactions in the DFTB and MD parts of the loop, we demonstrate numerically that the interaction between a classical point charge $Q$ and a molecule in a DFTB simulation is equivalent to the Coulomb interaction between $Q$ and the set of Mulliken charges $\{q_\alpha\}$ associated with those atoms, which inform the atomic charges in our MD simulations.

We consider a classical point charge $Q = -1.00\, e$ approaching the centre of a benzene molecule. DFTB simulations are carried out according to the method outlined in the Computational details section, with the exception that the threshold for convergence on the SCC-DFTB solution is set at $1 \times 10^{-6}\, E_H$ and the Fermi temperature to 10 K, this ensures convergence of the total energy. At each adsorption height, $d$, the force is extracted from the total energies of the DFTB simulations using a five-point stencil finite differences approach based on a series of equidistant simulations,

$$F = \frac{E^{Tot}_{-2\Delta d} + E^{Tot}_{+\Delta d} - E^{Tot}_{-\Delta d} - E^{Tot}_{+2\Delta d}}{12\Delta d} \quad \quad 3$$

Here $\Delta d (= 0.01\, \text{Å})$ is the spacing between ion height in each simulation and $E^{Tot}_{\pm i\Delta d}$ is the DFTB total energy at the adsorption height $\pm i\Delta d$.

The Coulomb Force between the $Q$ and $\{q_\alpha\}$ is computed vectorially as the sum

$$F = \frac{Q}{4\pi\epsilon_0} \sum_\alpha \frac{q_\alpha}{|\mathbf{r}_{Qq}|^3} \mathbf{r}_{Qq} \quad \quad 4$$

where $\mathbf{r}_{Qq}$ is the vector which gives the distance between the charges.

| Method | 2.00 Å | 3.00 Å | 10.00 Å |
|---|---|---|---|
| DFTB Total Energy | 87.45 | -27.55 | -0.35 |
| Coulomb Force | 87.43 | -27.54 | -0.35 |

**Table 1:** The total force exerted on a classical point charge for different adsorptions heights, computed as the first derivative of the DFTB total energy (eq. 1) and as the Coulomb force between the point charge and Mulliken charges on the atoms in a benzene molecule (eq. 2). All values reported in kJ mol-1nm-1

In table 1 we compare the total Coulomb force exerted on the classical point charge for three different adsorption heights 2.00, 3.00 and 10.00 Å. To within 0.02 kJ mol-1nm-1 the SCC-DFTB and classical coulomb forces are the identical, implying that Mulliken charges are fully transferable between our adopted methods.

### C. QM/MD simulations of graphene-electrolyte interfaces

To investigate the structure of the graphene-electrolyte interface, we start by calculating the electrolyte local mass density across the simulation box. In order to compensate for the fact that our graphene flake is finite, a *restricted* density profile is calculated, this means that contributions to the density profile are limited to a 2 nm radius around the vector which passes directly through the centre of the flake (radius $\approx 2.5$nm) as shown in Figure 3(c). This calculation removes from the results the effect of the graphene edges where the water and ions arrange differently.

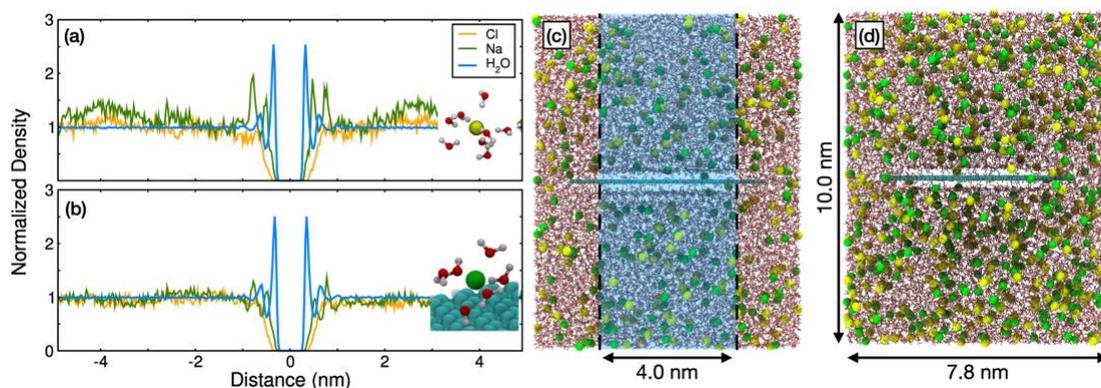

**Figure 3**: The restricted, normalised electrolyte densities in the direction parallel to the graphene surface normal for (a) 0.5 and (b) 1.0 M NaCl concentrations. Inset are examples of the ion hydration configurations for peaks closest to the graphene, (c-d) representative snapshots of the two different concentration simulation boxes, C, Na and Cl atoms are coloured cyan, green and yellow respectively.

Water in our simulations follows the typical SPC/E water-graphene structure,[19,58,66] with densities that are symmetric about the plane of the flake and two strong peaks at 3.4, 6.3 Å, and a third identifiable peak at 10.0 Å. It has been suggested that the increased order of the water molecules at this type of interface (electrolyte-graphene-electrolyte) is brought about by water-water interactions through the seemingly invisible graphene layer.[27] Our simulations have been carried out at two electrolyte concentrations, 0.5 and 1.0 M; in both cases the reduced density of $Cl^-$ close to the graphene indicates that the anion preferentially resides within the bulk water at distances greater than 1.0 nm from the flake. The weak attraction of the anion to the surface of the graphene flake is akin to the behaviour shown by smaller $F^-$ ions,[27] which are modelled based on a classical polarizable force field model and in agreement with the hybrid first principles/continuum simulations by Zhan *et al.*[32] The $Na^+$ ions have an increased density and demonstrates structuring at the interface. There are two strong peaks at 4.9 and 7.3 Å, which are directly in between the closest two water layers.

Previous work has shown that the presence and concentration of salts in the electrolyte seems to have very little effect on the overall structuring of water at the interface,[27,67] the QM/MD simulations confirm this. We also find that the relative positions of the water and $Na^+$ peaks in the density profile are unchanged moving from 0.5 to 1.0 M concentration. Yet, at higher concentrations, it appears that $Na^+$ ions are less likely to form structures at the interface since the peaks are significantly less intense with respect to bulk.

Further inspection of the simulations reveals additional emergent features in the QM/MD model, especially when compared with fully *ab-initio* molecular dynamics simulations of (comparatively smaller) graphene-water interfaces.[26] In addition to the layering of the water present in the density profile (Figure 3), the orientation of the molecular dipole moments relative to the plane of the graphene can provide extra information on fine structuring of the graphene-$H_2O$ interface.

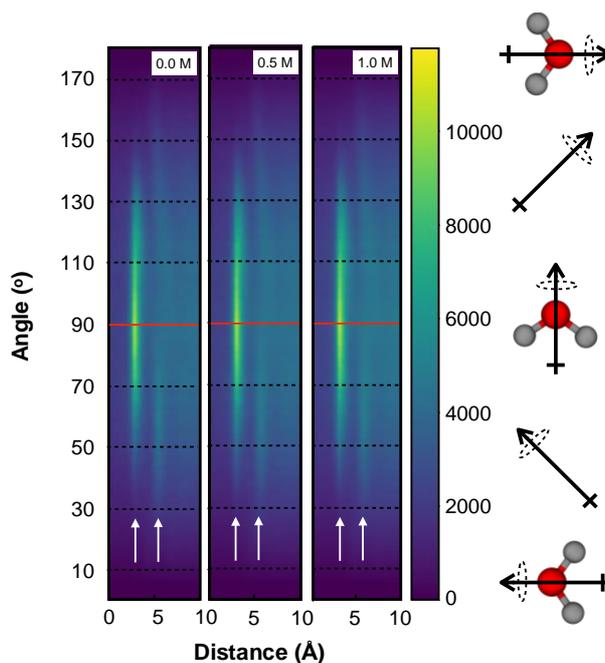

**Figure 4**: Two-dimensional histogram reporting the angular orientation of the water molecule molecular dipole moment relative to the graphene surface for three different salt concentrations. As a guide to the eye, white arrows indicate structuring and black arrows denote the orientation of the dipole for a given angle.

Figure 4 reports two-dimensional histograms, showing the distribution of molecular dipole moment angles relative to the graphene, as a function of vertical distance from the graphene flake for three different concentrations including pure water (e.g. zero salt concentration). As depicted in Figure 4, an angle of 90° corresponds to parallel alignment of the water dipole moment and the surface. Acute angles are indicative of a tilting of the negative end of the dipole (O atoms) towards the surface, conversely obtuse angles correspond to a tilting of the positive end of the dipole (H atoms) towards the surface. In our simulations the molecular dipole moments within the first layer of water molecules are aligned parallel to the graphene surface. We observe dipoles within the range 70° and 110° and none oriented perpendicular to the surface. The same orientation of molecular dipole moments has also been observed in modelling based on *ab-initio* molecular dynamics: (70° to 110° ),[24] (60° to 100°)[26] and polarizable force-field molecular dynamics: (65° to 120°),[27] and experimental measurements of water at hydrophobic interfaces.[68] This strict ordering of the molecular dipoles in the first layer induces a looser order in the second, with dipoles falling in the range 30° to 150°. Beyond 1 nm the structure induced by the graphene flake is diminished and all dipole-surface angles are distributed homogeneously.

One of the key findings arising from the use of the QM/MD loop is that, at concentrations equal to and below 1.0 M – and in the absence of any external bias – both the $Na^+$ or $Cl^-$ ions are more likely to remain hydrated when adsorbed on the graphene surface. This result can be understood by analysing the pairwise graphene-ion Coulomb and Lennard-Jones contributions to the energy. To this end, we performed simulations where one test ion (either $Na^+$ or $Cl^-$) is frozen at fixed heights above the centre of the graphene flake. In these calculations, for each height the simulation was run for 10 ns, with statistics collected after 5 ns every 5 ps. The resulting average Coulomb energies (which vary with the graphene polarization) and Lennard-Jones energies are plotted in Figure 5 for 3 different salt concentrations.

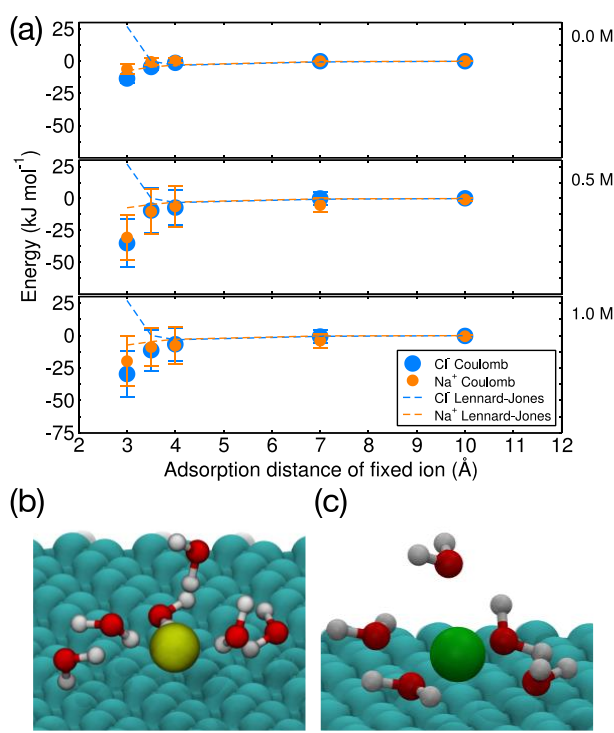

**Figure 5**: (a) Plot showing averaged Coulomb (dots) and Lennard-Jones (dashed lines) energies between a single frozen ion and the graphene flake at difference adsorption distances and at different molar concentrations of electrolyte. The partially solvated configurations for the (b) Cl and (c) Na ion at 3.0 Å are shown.

For distances equal to and greater than 4.0 Å, the short-range Coulomb and Lennard-Jones energies are approximately equal to zero, independent of the specific ion type or of the concentration of the solution. Closer to the graphene flake, the Lennard-Jones interaction plays a different role depending on the chemical species of the test ion. At 3.0 Å the graphene-$Cl^-$ Lennard-Jones energy is strongly repulsive, whilst the graphene-$Na^+$ interaction is weakly attractive. At this distance the short-range Coulomb interaction between $Cl^-$ and the polarised graphene is attractive, balancing the Lennard-Jones contribution. For each of the considered concentrations, within the first standard deviation, the net graphene-$Cl^-$ non-bonded interaction is neither attractive nor repulsive and therefore we cannot reasonably expect sustained adsorption of the ion at the surface. Interestingly, we observe that the strength of the graphene-ion Coulomb interaction is dependent on the concentration of the solution, which suggests that the magnitude of the graphene polarization is also linked to the number of ions present. At extreme dilution (0.0M), the short-range graphene-ion Coulomb energy is only very weakly attractive; but it becomes strongly attractive for higher concentrations (0.5 and 1.0M). This result indicates that the graphene flake is only weakly polarized by the water molecules alone, and the increased polarization is due to presence of other ions close to the interface.

The overall non-bonded interaction between $Na^+$ and graphene is also strongly attractive ($\approx -30$ kJ mol), and on this basis we could anticipate adsorption on the surface. Yet, as described above, the non-constrained QM/MD simulations (without fixed ions) at 0.5 and 1.0 M salt concentrations indicate $Na^+$ ions are more likely to be adsorbed at a distance from the graphene at which they retain their full solvation shell.

This phenomenon can be understood by considering the change in the Gibbs free energy during the dehydration of the ion ($-\Delta G_{\text{hydration}}$), which was computed for our model by Cheatham *et al* (369.9 kJ mol-1).[59] This energy, calculated by thermodynamic integration, corresponds to the removal of all of water molecules from the ion solvation shells. By calculating the ion-O radial distribution function (RDF) $g_{\text{NaO}}(\mathbf{r})$ (solid lines, Figure 6), we were able to compute the water molecule coordination number of the Na+ ion when frozen at a distance of 3.0 Å above the graphene surface as the integral of the first RDF peak (dashed lines, Figure 6). For all concentrations considered, the coordination of the ion is approximately 4.8 water molecules as depicted in Figure 5(c), which is in agreement with partially dehydrated-adsorbed configurations modelled by Williams *et al*.[14] However, looking at the density distributions obtained from the unconstrained simulations (Figure 3a and 3b) we observe that the most likely distance at which the cation is adsorbed is at 4 Å from the surface. At this distance the non-bonded interactions (Figure 5a) between Na+ and graphene are very weak justifying the modest cation's adsorption for distance shorter than this value. This result also indicates that even if stabilizing charge transfer interactions between the ion and graphene could occur, the (unconstrained) ions would have to adsorb much closer to the surface than their solvation spheres would allow in order for the transfer to take place. In fact, this conclusion is strengthened by available experimental measurements of graphite capacitance (which correlates with ion adsorption) in different group 1 metal solutions.[34] The capacitance increases as electrolyte solutions move down the group 1 alkali metals[34] – incidentally as the free energy of cation solvation is shown to decrease.[59]

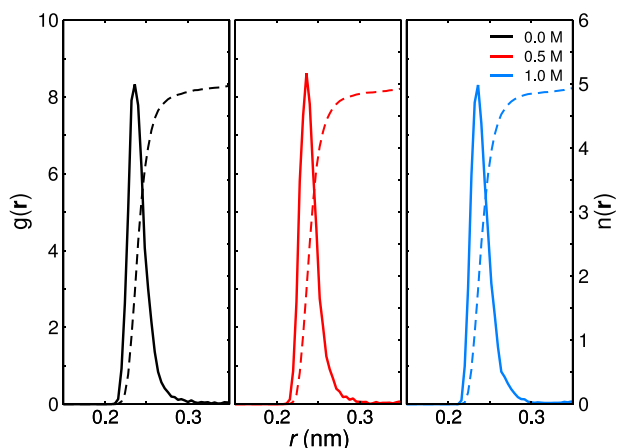

**Figure 6:** The first peak in the computed Na-O radial distribution function g($\mathbf{r}$) (solid lines) and cumulative integral, used to extrapolate the average coordination number n($\mathbf{r}$), (dotted lines) for systems with Na+ fixed at 3.0 Å above the graphene flake at different electrolyte concentrations.

The mild adsorption of cations and, even milder, anions on the non-charged surface of graphene, agrees with the recent results reported by from Zhan et al[32] using RISM theory. Our simulations for uncharged graphene surfaces confirm also that the ions density distribution close to the graphene sheet on opposing side of the flake is independent as found by Pykal et al[27] using polarizable force fields. In Figure 7b, the top panel shows the ions and water density profiles obtained for the simulation with the infinite sheet. These are very similar to those obtained for the flake simulation at 1M electrolyte (Figure 3b) with the peaks in the density profile near the surface found at the same distances. This result indicates that the structuring of the ions in proximity of the flake is not affected by the presence of another electrolyte solution on opposite side of the flake in agreement with Pykal. et al.27

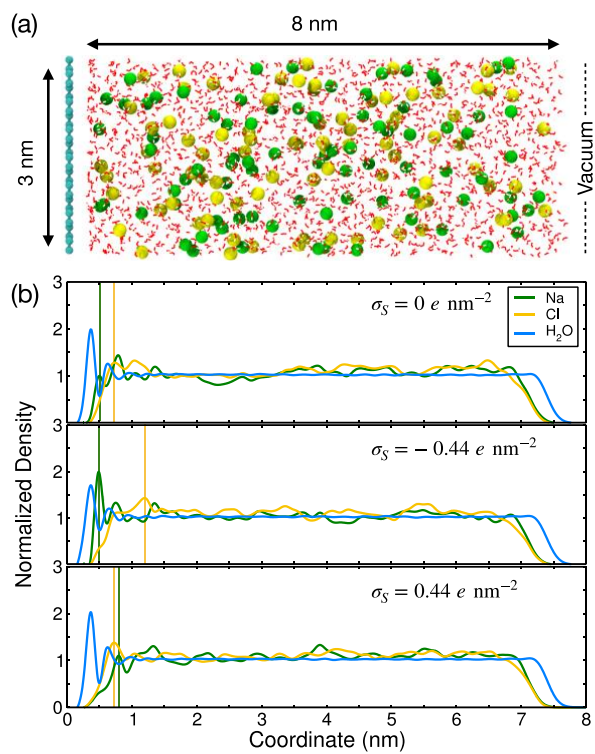

**Figure 7**. (a) Side view of the Graphene Electrolyte Vacuum interface investigated with the dimensions of the system marked, note that although not shown, 8 nm of vacuum is also applied. The C atoms in the graphene are coloured cyan, Na and Cl are green and yellow. (b) Corresponding plots of the electrolyte number densities normalised to bulk values for the considered surface charge densities ($\sigma_S$). As a guide to the eye, vertical lines mark the first peak in the Na and Cl densities.

The distance from the graphene surface at which the ions are adsorbed is however in disagreement with our own previous results using DFT with the conductor-like polarizable continuum model (PCM)14 and other DFT calculations performed with the ions in vacuum.21,69 As far as our previous work is concerned, the shorter ion-graphene adsorption distance that we observed is likely to be due to the use of the PCM, which implicitly describes the contribution of the solvent. In this case the continuum description of the solvent allows the ions to be smoothly adsorbed onto the surface by simulating a continuous dehydration process as opposed to the sharp removal of discrete number of water molecules from the first and second solvation shells. The result is an underestimation the dehydration free energy cost. For the DFT calculations performed in vacuum on finite size flake, the adsorption energy is affected by presence of a molecular quadrupole moment in the plane of the graphene flake which results in the over-binding of cations, and under-binding of anions, when compared with semi-infinite graphene models.52 The short ions adsorption distance predicted by Williams et al.14 is evident also if we compare the ions and water density profiles of Figure 7b with those from Dočkal et al.70 Here it is worth mentioning that there are several differences between ours and Dočkal's simulation set-up, most importantly different electrolyte concentrations and simulation box sizes which might affect the results.

The importance of the new multiscale procedure for modelling graphitic surfaces in contact with an electrolyte becomes evident monitoring the change in the polarization of the graphene electrode through the C atom charges. As part of the electronic supporting information we provide movies which show the real-time evolution of the atomic charges for a neutral, negative and positive graphene electrode. These movies can be correlated to the top, middle and bottom panels of Figure 7b to better appreciate the effect that charging the graphene sheet has on the ions and water density in proximity of the surface. In particular, it is interesting to notice that, at this magnitude of surface charge ($-0.44$ $e$ nm$^{-2}$), cations do not adsorb closer to the surface, rather the intensity of the peak at 0.49 Å, corresponding to the number cations adsorbed, increases (with respect to the non-charged system). For the anions instead, a clear adsorption peak at 0.72nm appears when the surface is positively charged, while there is no discernible adsorption for the non-charged case (Figure 7b).

Finally, the results obtained from the charged semi-infinite sheet setups can be used to evaluate how much the electrostatic potential of the electrode atoms changes during the simulations. This can be considered constant for an ideal conductor (if one ignores its underlying atomic structure) and this approximation may hold for atomistic simulations of metallic surfaces (this is the basic assumption of the Constant Potential Method[71–74]), but it is not immediately clear that it is valid in the case of graphene, which (*i*) is semimetallic under idealized conditions (*ii*) semiconducting in nanoribbons and quantum dots, and (*iii*) reportedly semiconducting under certain aqueous conditions.[75,76] Moreover, when considering the electronic structure of a material (not necessarily metallic), the connection with an electrode sets the *Fermi level* to a constant and, for sufficiently large systems, the net charge density and not the electrostatic potential. We can therefore use our QM/MD calculations to measure the deviation from this assumption and quantify the importance of incorporating the actual electronic structure of graphene rather than considering it as an ideal conductor. We thus estimate the electrostatic potential (calculated as the derivative with respect to the atomic charge of the total pairwise Coulomb energy for that atom), for each of the electrode atom in the two set-ups where the graphene sheet is charged. We notice (Figure 8) that the atomic electrostatic potential is not constant across the electrode and fluctuates between +/–1V and 0–2V for a graphene sheet charged positively and negatively respectively for the excess charge applied here (+/– 4$e$).

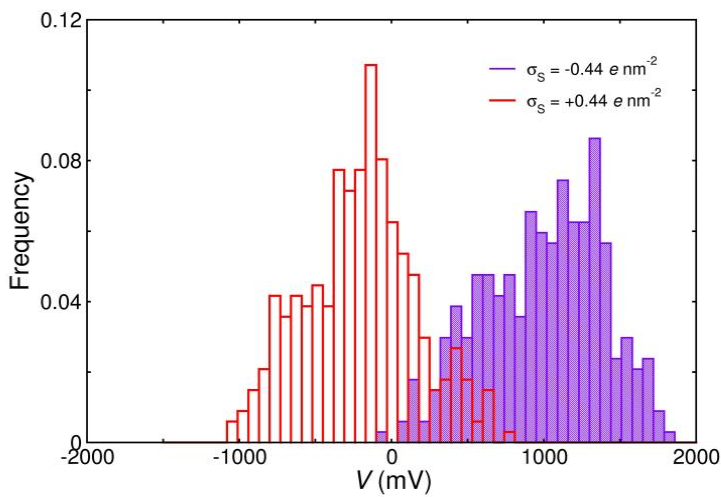

**Figure 8**. Histogram of the values of the electrostatic potential, *V*, calculated for each electrode C atom for one randomly selected snapshot of the QM/MD trajectories with the graphene positively (red) and negatively charged (blue).

## 1. Conclusions

In this work we propose a novel method to couple DFTB with classical MD to investigate the non-faradic interfacial adsorption processes occurring at carbon-based solid/liquid interfaces. Unlike other DFTB/MM methods the primary focus here is on the spatial- and time-resolution of the classical component of the system (i.e. the electrolyte solutions). This QM/MD approach captures the long-range electrostatic effects that originate from the interface allowing the simulation of large systems for long times and simultaneously models (*i*) the quantum mechanically informed surface polarizability, (*ii*) the electrolyte solutions including its rearrangement at the surface and (*iii*) the dynamical behaviour of fully solvated ions.

The method was applied to charge neutral graphene flakes, 5 nm in diameter, immersed in NaCl solutions of different concentrations for which the surface charge redistribution brought about by the proximity of the ions is large and delocalized over several carbon atoms. Density profiles and molecular dipole moment analysis suggests that water and $Na_+$ ions structure at the interface whereas $Cl_-$ ions remain mainly in bulk solution. We find that the electrolyte concentration does not have any effects on the distance from the surface at which the ions are adsorbed but does impact on the number of ions adsorbed: At high electrolyte concentrations ions are less adsorbed. Analysis on the energies of various fixed-ion configurations revealed that water alone does not polarize the surface and that the lack of overall adsorption can be rationalized as the adsorbed ion graphene configuration being thermodynamically disfavoured in comparison to the solvated ion configuration. This leads to the conclusion that charge transfer between the graphene and ions is unlikely to occur since the ions never get close enough to the surface for the transfer to take place and that the ions adsorption is, in the set-up explored here, only driven by the thermodynamics of the solution.

We also used our method to simulate a non-charged and charged infinite graphene sheets in contact with electrolyte on one side and vacuum on the other. The electrolyte properties of the former closely followed that obtained from graphene flake simulations. Upon charging, the intensity of ion adsorption is modulated according to whether the graphene is cathodic or anodic, however, the position of the adsorbed ions relative to the surface remains unchanged. Finally calculating the electrostatic potential across the electrode, we observe that the graphene flake (in this configuration and with this excess charge) does not show ideal metallic behaviour.

It is important to notice that, as in any classical simulations, the choice of the classical force field[77] (especially of the water model) in the MD step may affect subtly the ions adsorption distance and care needs to be take in the choice of the classical models.

In principle, this method can be applied to a wide array of polarisable materials, electrolyte solutions and their interfaces and different experimental set-ups including capacitors although in this case the extra challenge of simultaneously charging both electrodes during the quantum mechanical step of the loop needs to be solved. This method opens up the possibility for the simulation of electrified interfaces where the redistribution of excess charge on the electrode can be explicitly taken into account and any changes in the electronic properties of the surface can be investigated.

**Acknowledgements**
JDE and PC thank the European Union's Horizon 2020 research and innovation programme project VIMMP under grant agreement N$_o$ 760907. AT thanks the support of EPRSC. The authors would like to acknowledge the assistance given by Research IT, and the use of The HPC Pool funded by the Research Lifecycle Programme at The University of Manchester. The authors would like to acknowledge Dr Chris Williams for critical reading of the manuscript and constructive discussions.

**Supporting Information**
Plots showing the convergence of DFTB Mulliken charges, plot showing the change in the average computed DFTB Mulliken charge over a classical MD trajectory with different time steps. Movies of the time evolution of surface charge for neutral, cathodic (negatively charged) and anodic (positively charged) surfaces.